\documentclass[runningheads]{llncs}

\usepackage[T1]{fontenc}

\usepackage{hyperref}
\usepackage{graphicx}
\usepackage{color}

\usepackage{subcaption}

\usepackage{booktabs}

\usepackage[most]{tcolorbox}
\usepackage{spverbatim}

\usepackage{algorithm}
\usepackage{algpseudocode}

\usepackage{fancyhdr}
\usepackage{lipsum}

\begin{document}

\title{ChatSpamDetector: Leveraging Large Language Models for Effective Phishing Email Detection}

\titlerunning{ChatSpamDetector}

\author{Takashi Koide \and
Naoki Fukushi \and
Hiroki Nakano \and
Daiki Chiba}
\authorrunning{T. Koide et al.}
\institute{NTT Security Holdings, Tokyo, Japan \email{\{takashi.koide,naoki.fukushi\}@global.ntt,\\hi.nakano.sec@gmail.com,daiki.chiba@ieee.org}}
\maketitle              %

\noindent\makebox[\textwidth][l]{%
  \raisebox{60mm}[0pt][0pt]{%
    \fbox{%
      \parbox{0.55\paperwidth}{%
        \footnotesize
        If you cite this paper, please use the following reference:\\
        Takashi Koide, Naoki Fukushi, Hiroki Nakano, and Daiki Chiba.: ChatSpamDetector: Leveraging Large Language Models for Effective Phishing Email Detection. In \emph{20th {EAI} International Conference on Security and Privacy in Communication Networks (SecureComm 2024)}, October 28--30, 2024, Dubai, United Arab Emirates.
      }%
    }%
  }%
}

\begin{abstract}
The proliferation of phishing sites and emails poses significant challenges to existing cybersecurity efforts. Despite advances in malicious email filters and email security protocols, problems with oversight and false positives persist. Users often struggle to understand why emails are flagged as potentially fraudulent, risking the possibility of missing important communications or mistakenly trusting deceptive phishing emails.

This study introduces \textsc{ChatSpamDetector}, a system that uses large language models (LLMs) to detect phishing emails. By converting email data into a prompt suitable for LLM analysis, the system provides a highly accurate determination of whether an email is phishing or not. Importantly, it offers detailed reasoning for its phishing determinations, assisting users in making informed decisions about how to handle suspicious emails. We conducted an evaluation using a comprehensive phishing email dataset and compared our system to several LLMs and baseline systems. We confirmed that our system using GPT-4 has superior detection capabilities with an accuracy of 99.70\%. Advanced contextual interpretation by LLMs enables the identification of various phishing tactics and impersonations, making them a potentially powerful tool in the fight against email-based phishing threats.

\keywords{Large language models  \and Phishing emails.}
\end{abstract}

\section{Introduction}

The damage caused by phishing sites is growing every year, with phishing emails being actively used by attackers as a longstanding method of luring users. Modern email services have incorporated sophisticated malicious email filters and authentication protocols to mitigate these threats by identifying suspicious senders and analyzing email content for potential phishing indicators~\cite{gmail-spam,outlook-spam}. However, despite these efforts, oversights and misclassifications of phishing emails persist.
Many email services may classify emails as potentially fraudulent without providing users with an explicit rationale for the classification decision. As a result, recipients cannot determine if an email is truly phishing and may accidentally miss an important message or unintentionally click on a malicious link.

In this study, we propose a system, \textsc{ChatSpamDetector}, that uses Large Language Models (LLMs), such as ChatGPT, to effectively detect phishing emails. By generating prompts to LLMs based on email body and header information, our system leverages the powerful analytical capabilities of LLMs to detect phishing emails. This approach not only enables highly accurate identification of phishing emails, but also provides detailed explanations for the classification decision. 
While previous research~\cite{Heiding2024DevisingAD} has explored the use of LLMs to automatically generate phishing emails and analyze their intent, the ability of LLMs to distinguish between phishing and legitimate emails has not been clearly demonstrated.
Our system is the first study to evaluate the ability of LLMs to detect phishing emails.

By automatically generating reports detailing the classification results, our system helps users make informed decisions for subsequent actions, such as discarding suspicious emails or trusting legitimate ones. To evaluate the performance of our proposed system, we prepared a dataset containing both phishing emails and legitimate emails, and compared the detection accuracy with several LLMs and baseline systems. The results showed that our system using GPT-4 achieved an accuracy of 99.70\%, outperforming other models and baseline systems. Using our unique prompts, this system increases the true positive rate compared to simple prompts.

Through detailed analysis of LLM responses, we confirmed the ability of LLMs to extract key indicators from the headers and body of emails, prioritize them, and generate accurate responses, confirming their effectiveness in phishing detection.
We have observed that the proposed system accurately identifies sender spoofing and brand impersonation, along with various social engineering (SE) techniques (e.g., account anomaly notifications, cryptocurrency airdrops, virus infection alerts) contained in phishing emails. The system can also reveal the various techniques attackers use to evade malicious email filters.

Our system represents a new step in the fight against phishing emails, as it can be used to replace or complement the existing phishing detection features of email services, helping users to determine the suspiciousness of emails on their own. By providing users with the tools to make informed decisions about the legitimacy of emails, our proposed system enables individuals to protect themselves against the evolving phishing attack landscape.

The contributions of this paper are summarized below:
\begin{itemize}
\item We proposed \textsc{ChatSpamDetector}, a system that uses LLMs to detect phishing emails. This system not only achieves high accuracy in identifying phishing emails, but also provides users with informative explanations supported by concrete reasons.
\item We collected recent phishing emails and legitimate emails to create datasets, and conducted evaluation experiments on them. The results demonstrated that our proposed system using GPT-4 achieved 99.70\% accuracy, outperforming other models and baseline systems.
\item We performed a detailed analysis of LLM's responses, which shows LLM's sophisticated ability to extract important information from the header and body of an email, prioritize them, and output an accurate response based on a comprehensive analysis.
\end{itemize}

\section{Background}

To enhance defenses against phishing emails, many studies have been proposed that use machine learning-based methods to analyze and categorize emails. These methods statistically extract phishing-specific features from headers and bodies for binary (ham/spam) classification~\cite{DBLP:journals/corr/abs-2203-10408,DBLP:journals/cluster/MughaidAHTAE22,DBLP:conf/raid/NabeelASKWY21,DBLP:journals/sncs/Sonowal20}. In addition, methods using natural language processing to examine the text of phishing emails and interpret their context and intent have been explored\cite{DBLP:conf/qrs/CheLZYZY17,DBLP:journals/tbd/LiCWS22}. These methods range from simple frequency-based feature vector generation, such as TF-IDF~\cite{mo-messidi} and Bag-of-words~\cite{MoAbd}, to advanced deep learning models such as BERT (Bidirectional Encoder Representations from Transformers), which use semantic analysis to detect SE techniques~\cite{DBLP:conf/eurosp/LeeTYAHD21,DBLP:conf/codaspy/QachfarVM22,DBLP:conf/ccs/SakaVK22}.
While many of these machine learning-based methods demonstrate high performance on existing phishing email datasets, they primarily offer statistical detection results without providing specific rationales. Furthermore, models trained on limited datasets may not perform effectively against various types of new phishing emails in different languages.

Email services such as Gmail and Outlook have implemented sophisticated spam filters that use techniques such as blocklist methods and machine learning to accurately identify spam based on email content and the authenticity of the senders~\cite{gmail-spam,outlook-spam}. However, most services do not explain why an email is classified as spam, and legitimate emails may be mistakenly sorted into the spam folder, or spam emails may be missed and arrive in the inbox. Ultimately, the decision to trust and open an email is left to the user, who may not have enough information to make an informed decision.

Several email security protocols have been proposed to reduce phishing emails. SPF (Sender Policy Framework)~\cite{rfc4408} is an authentication protocol for verifying the sending domain of an email, allowing receiving servers to verify that an email comes from a trusted domain name. DKIM (DomainKeys Identified Mail)~\cite{rfc6376} uses signed headers to ensure that an email has not been tampered with and verifies the legitimacy of the sender. DMARC (Domain-based Message Authentication, Reporting, and Conformance)~\cite{rfc7489} integrates SPF and DKIM, allowing domain owners to set email authentication policies, prevent unauthorized sending, and receive reports. BIMI (Brand Indicators for Message Identification) is a standard that enables brands to display their logos in email inboxes, providing visual authentication and enhancing brand recognition~\cite{brand-indicators-for-message-identification-04,DBLP:conf/pam/YajimaCYM23}. While these security measures help reduce phishing emails, attackers have developed various techniques to circumvent these protections, such as display name and domain name spoofing.

\section{ChatSpamDetector}

\begin{figure}[!t]
    \centering
        \includegraphics[width=\linewidth]{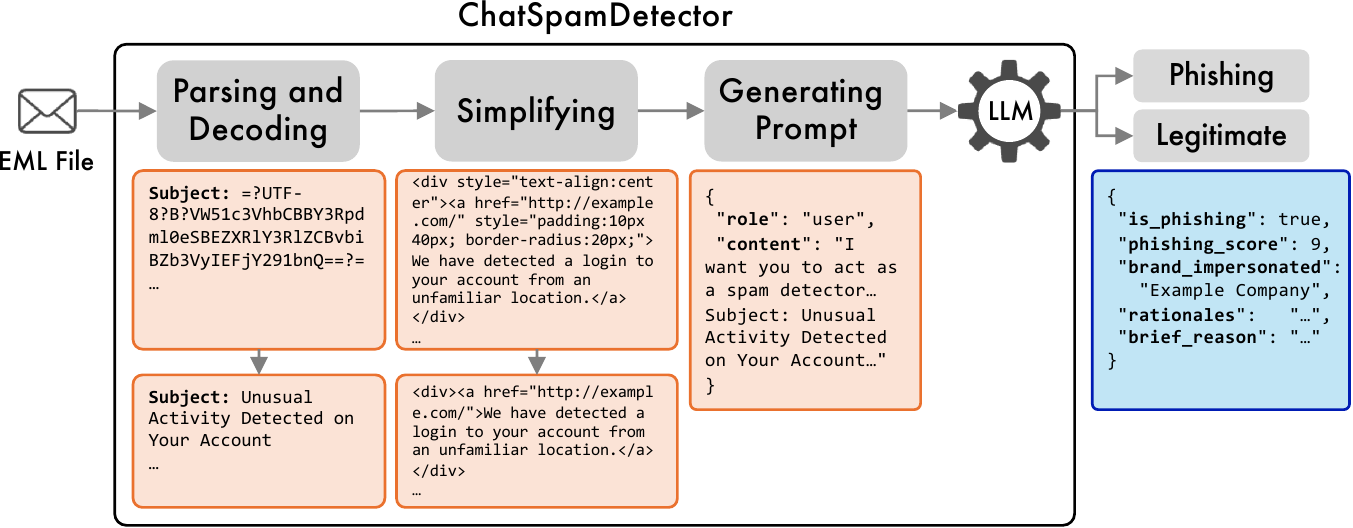}
    \caption{System Architecture of \textsc{ChatSpamDetector}}
    \label{fig:system}
\end{figure}

In this study, we present \textsc{ChatSpamDetector}, a system that uses LLMs to detect phishing emails. Our system converts emails into appropriate prompts for LLMs to analyze the entire email content. This enables advanced contextual analysis to detect complex SE techniques that cannot be detected by previous methods, including advanced deep learning techniques. Even for phishing emails that bypass email authentication, the system can identify brand impersonation based on the difference between the brand name in the email and its legitimate domain name. In addition, unlike spam filters embedded in email services, our system not only outputs the detection results, but also provides a detailed report, including the rationale behind its decisions. Our system accepts email data in \texttt{.eml} format and outputs a JSON report containing the phishing detection results. Targeted phishing emails include those that contain links to phishing sites designed to trick recipients into visiting them. 
The system architecture of \textsc{ChatSpamDetector} is shown in Figure~\ref{fig:system}.
The process of our system—email parsing and decoding, simplification, prompt generation, and request handling—is detailed below.

\subsection{Parsing and Decoding Email}
An \texttt{.eml} file represents an email message in text format that includes a header containing metadata and formatting information, such as sender and time stamp information, and a body containing the message and attached files. Since headers and bodies can be encoded in different character sets and encoding schemes, our system decodes them and reconstructs the email for easier interpretation by LLMs. 
For example, the email subject \texttt{Subject: Security Alert!} when encoded using the UTF-8 character set and Base64 encoding, appears as \texttt{Subject: =?UTF-8?B?U3ViamVjdDogU2VjdXJpdHkgQWxlcnQh?=}. Such strings, while generally interpretable by the advanced parsing capabilities of LLMs for short texts, increasingly pose a risk of failure as they lengthen.

To avoid unnecessary overhead in LLM analysis, header information such as DKIM and DMARC authentication signatures or custom headers starting with \texttt{X-} are removed. For experimental fairness, headers related to any spam filtering decisions made by email servers are excluded from our dataset. Attachments are handled by extracting filenames and discarding binary data.

\subsection{Simplifying Email}
Emails can be long or contain complex HTML structures. Given the token limit constraints of LLMs, it is essential to adjust the input data to not exceed these limits. Our system sets a preliminary token limit and simplifies the email body if this threshold is surpassed. In this paper, we set a token limit of 3,000 for the entire email text. This limit is based on the 4k token limit of Llama 2, taking into account the tokens consumed by the prompt template and responses. The simplification process is adaptable to different LLM specifications and includes steps for measuring token numbers, organizing multipart emails, and omitting unnecessary content in both \texttt{text/html} and \texttt{text/plain} formats. The specific process involves the following four steps.

\noindent\textbf{Measuring the Number of Tokens}
First, Our system counts the number of tokens in the emails using the tokenizer specific to each LLM. This step is necessary because the token count can vary depending on the tokenizer used by each LLM. If the token limit is exceeded, our system performs the following steps sequentially until the token count falls below the limit.

\noindent\textbf{Organizing Multi-part Email}
Multipart emails contain a mix of HTML and plain text messages. Messages in HTML format often contain phishing tactics designed to deceive users with their rich appearance. For example, the hyperlink displayed in a message may differ from the actual link specified in the \texttt{href} attribute. Our system prioritizes HTML messages when multiple message formats are present, and deletes other messages to save tokens.

\noindent\textbf{Omitting \texttt{text/html} Content}
We describe a method for omitting HTML format messages. The primary goal of phishing emails is to direct the viewer to a phishing site by clicking on hyperlinks. While retaining crucial information such as the \texttt{a} tag and text containing false information to deceive users, our system removes unnecessary HTML elements related to style or metadata. The removal process is as follows: Inspired by existing research on using LLMs to detect phishing sites~\cite{koide2023detecting}, we implemented a simplification of the HTML.
\begin{enumerate}
    \item Remove \texttt{commnet}, \texttt{styl}, \texttt{script} tags.
    \item Remove all attributes except for the major ones: \texttt{src}, \texttt{href}, \texttt{alt}, \texttt{title}, \texttt{name}, \texttt{id}, \texttt{class}.
    \item Remove tags with no text content.
    \item Unwrap certain tags: \texttt{font}, \texttt{strong}, \texttt{b}.
    \item Reduce the \texttt{src} attribute of \texttt{img} tags and the \texttt{href} attribute of \texttt{a} tags to 10 characters in the URL path, except for the domain name.
    \item If the token limit is exceeded, remove an HTML element from the center of the HTML until the number of tokens is below the limit.
\end{enumerate}

\noindent\textbf{Omitting \texttt{text/plain} Content}
It is common for the top of an email to contain important information that grabs the recipient's attention and facilitates a quick understanding of the email content. Similarly, the bottom of an email often contains the sender's signature or contact information. To preserve such important information, while reducing the number of tokens, our system removes the line from the middle of the message until the total token count falls below the limit.

\definecolor{promptcolor}{RGB}{150,75,0} %
\definecolor{responsecolor}{RGB}{0,0,128} %
\newtcbtheorem[]{prompt}{Prompt Template}%
{
  colback=promptcolor!10!white,
  colframe=promptcolor,
  coltext=black,
  boxsep=0pt,
  fonttitle=\bfseries,
  fontupper=\ttfamily,
  sharp corners,
}{prompt}

\begin{figure}[!t]
\begin{prompt}[label=prompttemp1]{Normal Prompt}{prompttemp1}
I want you to act as a spam detector to determine whether a given email is a phishing email or a legitimate email. Your analysis should be thorough and evidence-based. Phishing emails often impersonate legitimate brands and use social engineering techniques to deceive users. These techniques include, but are not limited to: fake rewards, fake warnings about account problems, and creating a sense of urgency or interest. Spoofing the sender address and embedding deceptive HTML links are also common tactics.\\ 
Analyze the email by following these steps:\\
1. Identify any impersonation of well-known brands.\\
2. Examine the email header for spoofing signs, such as discrepancies in the sender name or email address. \\Evaluate the subject line for typical phishing characteristics (e.g., urgency, promise of reward). Note that the To address has been replaced with a dummy address.\\
3. Analyze the email body for social engineering tactics designed to induce clicks on hyperlinks. Inspect URLs to determine if they are misleading or lead to suspicious websites.\\
4. Provide a comprehensive evaluation of the email, highlighting specific elements that support your conclusion. Include a detailed explanation of any phishing or legitimacy indicators found in the email.\\
5. Summarize your findings and provide your final verdict on the legitimacy of the email, supported by the evidence you gathered.\\

Email:\\
\verb|```<Insert email text data here>'''|
\end{prompt}
\end{figure}

\subsection{Generating Prompt}
\label{sec:generatingprompt}

LLMs can improve the accuracy of their responses by customizing input prompts and it is known as prompt engineering~\cite{white2023prompt}. Using prompt engineering techniques, we created Prompt Template~\ref{prompttemp1} to analyze an email assigned to the template. The specific process to instruct LLMs is following.

\noindent\textbf{Assigning Roles} Assigning specific roles tailored to the task to the LLM is known to enhance the model's response. We assign the LLM the persona of a \textit{spam detector}, guiding it to objectively analyze emails.

\noindent\textbf{Setting Subtasks} Chain-of-thought prompting~\cite{KojimaGRMI22,Wei0SBIXCLZ22} improves the reasoning abilities of LLMs by presenting them with problem-solving examples and their associated steps. This helps LLMs understand the logic of the problem and produce accurate, understandable responses.
Based on chain-of-thought prompting and previous research~\cite{koide2023detecting}, we have developed a specific prompt for detecting phishing emails. By dividing the task into five subtasks and analyzing them sequentially, we guide the model through the correct analytical process. Specifically, the model is instructed to examine the textual content and header fields of the email to identify any psychological manipulation or deception indicative of social engineering. In addition, the model is prompted to present evidence of phishing, accompanied by detailed explanations to support its conclusions.

\noindent\textbf{Providing Examples of SE Techniques} Due to the ambiguous definition of phishing emails, LLMs may make different assumptions in their analysis. Our system provides LLMs with examples of common SE techniques and enables them to detect such emails. This is intended to establish phishing detection criteria and improve the consistency and reliability of LLM analysis.

\subsection{Requesting}
Our system systematically retrieves detection results from LLMs by defining the response format through Function Calling, a feature that provides the necessary arguments (in JSON format) for calling predefined functions. The function settings are as shown in Table~\ref{table:function_calling}. This function is named \texttt{print\_phishing\_result}, with a description of ``Outputs whether a given email is a phishing email or a legitimate email.''
This allows LLMs to generate a structured output, including the determination of whether an email is phishing (\texttt{is\_phishing}), a phishing confidence score (\texttt{phishing\_score}), the impersonated brand (\texttt{brand\_inpersonated}), detailed rationales (\texttt{rationales}), and a brief summary of the decision reasons (\texttt{brief\_reason}). 
Outputting the reason for the decision helps the user make an informed decision about whether to discard or open the email.
Our system inserts the shortened email into a prompt template, submits it to an LLM, and interprets the JSON data from the response. Note that LLMs have configurable parameters that can affect the output, such as the variety and randomness of the generated text. For example, ChatGPT allows adjustments to parameters such as \texttt{top\_p} and \texttt{temperature}. This study uses the default values for all parameters in each LLM.

\begin{table}[!t]
\centering
\caption{Properties of Function Calling.}
\label{table:function_calling}
\begin{tabular}{llp{8cm}}
\toprule
\textbf{Property Name} & \textbf{Type} & \textbf{Description} \\
\hline
is\_phishing & boolean & A boolean value indicating whether the email is phishing (true) or legitimate (false). \\
\hline
phishing\_score & number & Phishing risk confidence score as an integer on a scale from 0 to 10. \\
\hline
brand\_impersonated & string & Brand name associated with the email, if applicable. \\
\hline
rationales & string & Detailed rationales for the determination, up to 500 words. \\
\hline
brief\_reason & string & Brief reason for the determination. \\
\bottomrule
\end{tabular}
\end{table}

\section{Dataset}
To evaluate the performance of our proposed system, we prepared a dataset consisting of phishing emails. The goal of the experiment is to verify the system's ability to accurately distinguish between phishing and legitimate emails. We describe the process used to collect these emails.

\noindent\textbf{Phishing Email}
Typical datasets used for spam detection include Enron email dataset~\cite{enron}, SpamAssassin public mail corpus~\cite{spamassasincorpus}, and the TREC Public Spam Corpus~\cite{treccorpus}. However, these datasets are quite outdated and include emails from before the implementation of email security protocols such as SPF, DKIM, and DMARC. Consequently, these datasets may not include the latest phishing techniques or vulnerable brands, including those related to social media and cryptocurrency scams. To address this, we used phishing\_pot~\cite{phishingpot} as our source, which shares the latest phishing emails, covering different languages and brands. This repository captures phishing emails from honeypots and is updated regularly to provide \texttt{.eml} data. From the emails collected between August 2022 and October 2023, we included 1,010 phishing emails in our dataset, excluding emails without URLs in the body; this essentially includes all emails that lead to phishing sites. These emails spanned 19 languages, with 627 in English, 128 in Portuguese, 95 in German, and 79 in Dutch, targeting 193 brands. Examples of brands include online retail (Amazon, Best Buy, Otto, Walmart), technology and software (Apple, Adobe, Github, Microsoft), finance and banking (American Express, Bradesco, Paypal, MetaMask), social media (Facebook, Pinterest, WhatsApp), and shipping and logistics (UPS, Chronopost, Correios).

\noindent\textbf{Legitimate Email}
We selected legitimate emails from the CSDMC SPAM corpus~\cite{csdmc}, which has been commonly used in previous studies and phishing detection competitions. This dataset contains 2,949 ``ham'' emails, from which we randomly sampled 1,000, roughly the same number as phishing emails, after excluding those with sender addresses anonymized to domain names such \urlstyle{tt}\url{example[.]com}. The final dataset comprised emails in 12 languages, including 895 in English, 33 in Dutch, 30 in French, and 25 in Romanian.

\section{Evaluation}

To assess the accuracy of \textsc{ChatSpamDetector}, we conducted experiments using the dataset with various LLMs and baseline systems.

\subsection{Model Selection}

\begin{figure}[!t]
\begin{prompt}[label=prompttemp2]{Simple Prompt}{prompttemp2}
Determine whether a given email is a phishing email or a legitimate email.\\
\\
Email:\\
\verb|```<Insert email text data here>'''|
\end{prompt}
\end{figure}

\begin{figure}[!t]
\begin{prompt}[label=prompttemp3]{Additional Prompt for LLama 2}{prompttemp3}
\begin{tcolorbox}[left=0pt,right=0pt,top=0pt,bottom=0pt,boxrule=0.5pt]
Note: Here is the same as Prompt Template~\ref{prompttemp1}, except for the email insertion part.
\end{tcolorbox}
6. Your output should be JSON-formatted text with the following keys:\\
- is\_phishing: a boolean value indicating whether the email is phishing (true) or legitimate (false)\\
- phishing\_score: phishing risk confidence score as an integer on a scale from 0 to 10\\
- brand\_impersonated: brand name associated with the email, if applicable\\
- rationales: detailed rationales for the determination, up to 500 words\\
- brief\_reason: brief reason for the determination\\
\\
Email:\\
\verb|```<Insert email text data here>'''|
\end{prompt}
\end{figure}

In our experiments, we assessed four models: the GPT-3.5-Turbo (referred to as GPT-3.5), GPT-4, Llama2-70B (referred to as Llama 2), and Gemini Pro. The GPT-4 and GPT-3.5 models were accessed via the Azure OpenAI Service~\cite{azureopenai} using the \texttt{gpt-4-0613} and \texttt{gpt-35-turbo-0613} versions, respectively. We deployed Llama 2 on Azure Machine Learning using the \texttt{NC96ads\_A100\_v4} instance with four NVIDIA A100 GPUs. We accessed Gemini Pro via the Google Cloud API using the \texttt{gemini-pro@001} model~\cite{geminiapi}. To analyze the effect of variations in prompts on detection accuracy, we created a simple prompt as shown in Prompt Template~\ref{prompttemp2}. In this case, we only set the \texttt{is\_phishing} parameter from Table~\ref{table:function_calling} for Function Calling. Since Llama 2 does not support Function Calling, we created the Prompt Template~\ref{prompttemp3} to get responses including JSON format. 
The simple prompt for Llama 2 is this template with all keys removed except for \texttt{is\_phishing}.
Some Llama 2 responses did not contain a valid JSON format string, so we heuristically parsed them to extract the true/false phishing detection results.
We used Function Calling for all the other models.

\subsection{Results}
\label{section:results}

\begin{table}[!t]
\centering
\caption{Results of \textsc{ChatSpamDetector} and baseline systems.}
\label{tab:evaluation_results}
\small
\scalebox{0.97}[1.0]{
\begin{tabular}{lll|rrrr|rrr}
\toprule
        System & Model & Prompt & TP & FP & TN & FN & Precision & Recall & Accuracy \\ 
        \midrule
        ChatSpamDetector & GPT-4 & Normal & 1,007 & 3 & 997 & 3 & \textbf{99.70\%} & \textbf{99.70\%} & \textbf{99.70\%} \\ 
        ~ & ~ & Simple & 1,001 & 4 & 996 & 9 & 99.60\% & 99.11\% & 99.35\% \\ 
        ~ & GPT-3.5 & Normal & 980 & 32 & 968 & 30 & 96.84\% & 97.03\% & 96.92\% \\ 
        ~ & ~ & Simple & 697 & 6 & 994 & 313 & 99.15\% & 69.01\% & 84.13\% \\ 
        ~ & LLama 2 & Normal & 950 & 361 & 639 & 60 & 72.46\% & 94.06\% & 79.05\% \\ 
        ~ & ~ & Simple & 790 & 9 & 991 & 220 & 98.87\% & 78.22\% & 88.61\% \\ 
        ~ & Gemini Pro & Normal & 991 & 21 & 979 & 19 & 97.92\% & 98.12\% & 98.01\% \\ 
        ~ & ~ & Simple & 977 & 6 & 994 & 33 & 99.39\% & 96.73\% & 98.06\% \\ 
        \midrule
        Baseline A~\cite{dbsheta} & - & - & 580 & 374 & 626 & 430 & 60.80\% & 57.43\% & 60.00\% \\ 
        Baseline B~\cite{mo-messidi} & - & - & 564 & 413 & 587 & 446 & 57.73\% & 55.84\% & 57.26\% \\ 
        Baseline C~\cite{MoAbd} & - & - & 923 & 827 & 173 & 87 & 52.74\% & 91.39\% & 54.53\% \\ 
        Baseline D~\cite{DBLP:journals/corr/abs-2203-10408} & - & - & 941 & 208 & 792 & 69 & 81.90\% & 93.17\% & 86.22\% \\ 
\bottomrule
\end{tabular}
}
\end{table}

Here, we describe the experimental results, which are shown in Table~\ref{tab:evaluation_results}.

\subsubsection{Summary}
First, we measured recall, also known as the true positive rate (\(= \frac{TP}{TP + FN} \)), where TP represents true positives and FN represents false negatives. GPT-4 with the normal prompt was the highest at 99.70\%, outperforming GPT-3.5 (97.03\%), Llama2 (94.06\%) and Gemini Pro (98.12\%).
Comparisons of results between simple and normal prompts show an improvement in recall across all models, indicating that simple prompts tend to be significantly biased towards negative (legitimate) classifications. In fact, the false positives were significantly lower with simple prompts across all models, with 4, 6, 9, and 6 for GPT-4, GPT-3.5, Llama 2, and Gemini Pro, respectively. The use of normal prompts improved the models' ability to accurately identify phishing emails while reducing false positives. 
This improvement was particularly pronounced for the less performing models, with GPT-3.5 and Llama2 having recalls increased by 28.02\% and 15.84\%, respectively. 
However, the normal prompt for Llama 2 caused an increase in false positives (FPs) in contrast to the simple prompts, leading to a 26.41\% reduction in precision (\(= \frac{TP}{TP + FP}\)). 
When evaluating precision and accuracy (\(= \frac{TP + TN}{TP + FP + TN + FN}\), where TN represents true negatives), both metrics for GPT-4 using normal prompts were exceptionally high at 99.70\%, demonstrating its superiority over the other models.

\begin{figure}[!t]
  \centering
  \begin{subfigure}[b]{0.45\linewidth}
    \includegraphics[width=\linewidth]{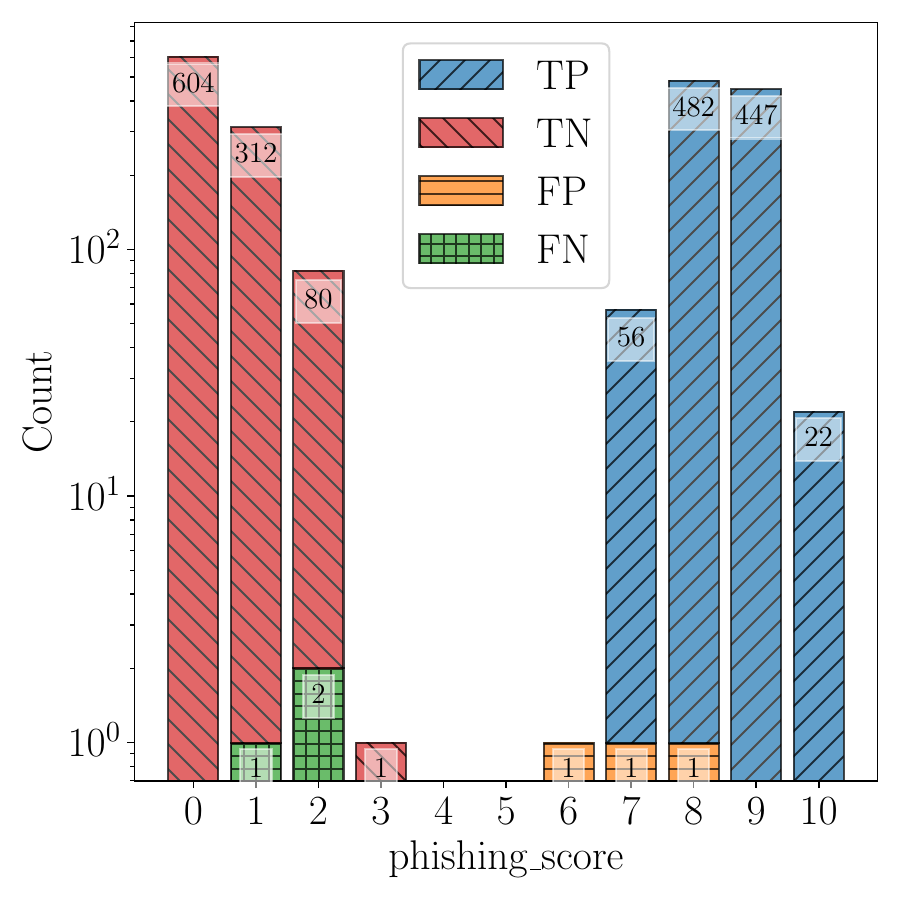}
    \caption{GPT-4}
    \label{fig:phishing_score_gpt4}
  \end{subfigure}
  \hfill
  \begin{subfigure}[b]{0.45\linewidth}
    \includegraphics[width=\linewidth]{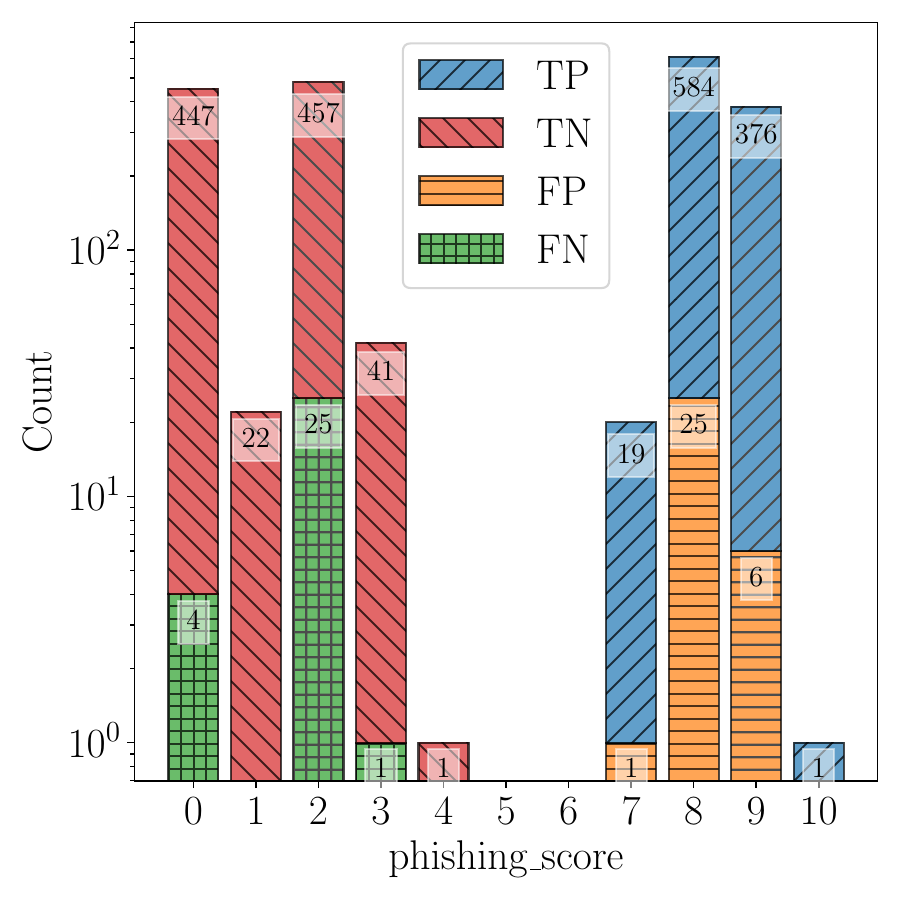}
    \caption{GPT-3.5}
    \label{fig:phishing_score_gpt3}
  \end{subfigure}
  \caption{Comparison of the \texttt{phishing\_score} for GPT-4 and GPT-3.5}
  \label{fig:phishing_score}
\end{figure}

\subsubsection{Phishing Score}
Figures~\ref{fig:phishing_score_gpt4} and~\ref{fig:phishing_score_gpt3} show the distribution of the \texttt{phishing\_score} when using GPT-4 and GPT-3.5 respectively, categorized into TP, TN, FP and FN with stacked bars. Overall, emails classified as phishing had higher scores, while those classified as legitimate had lower scores. For GPT-4, the scores for emails classified as legitimate (negative) ranged from 1 to 4, with TN scoring most frequently 0 (604 cases), indicating a high level of confidence in correctly identifying legitimate emails. On the other hand, scores for emails identified as phishing (positive) were distributed between 6 and 10, with peaks at 8 and 9. GPT-3.5 had high scores such as 8 and 9 for FP and low scores such as 0 for FN, while GPT-4's FP and FN were more neutrally distributed around 6, 7, 8 for FP and 1, 2 for FN, indicating that GPT-4 is less likely to make mistakes when providing higher confidence responses.

\subsubsection{API Request Latency}
We describe the response time when experimenting with the proposed system. The average time from sending the request to receiving the response was 12.53 seconds for GPT-4 and 2.50 seconds for GPT-3.5. Note that these results are subject to change based on various factors, such as the region of the Azure OpenAI Service and server congestion.

\subsubsection{Cost Analysis}
We did a cost analysis for the experiments using the ChatGPT API (GPT-4 and GPT-3.5). The cost per input was \$0.03 for GPT-4 and \$0.002 for GPT-3.5, while the cost per output was \$0.06 for GPT-4 and \$0.002 for GPT-3.5. Analyzing the 2,010 emails in our dataset, the cost of using GPT-4 with a normal prompt was \$265.80 and for GPT-3.5 was \$13.68, while using a simple prompt resulted in costs of \$249.93 for GPT-4 and \$12.62 for GPT-3.5.

\subsection{Comparison with Baseline Systems}

We experimented with four different types of baseline phishing email detection systems to compare the detection accuracy of our systems.
\\\noindent\textbf{Baseline A}~\cite{dbsheta} tokenizes the body of emails and uses a convolutional neural network (CNN) for binary classification to determine whether an email is phishing or not. As shown in the use case of the original code, we trained the model on the Enron dataset and then applied the same tokenization to our dataset for experimentation.
\\\noindent\textbf{Baseline B}~\cite{mo-messidi} extracts features from the email body using TF-IDF and classifies them using a Support Vector Machine (SVM). We used a model trained on the data included in this code and performed experiments on our dataset.
\\\noindent\textbf{Baseline C}~\cite{MoAbd} extracts features from the email body using a bag-of-words algorithm and classifies emails using a neural network. We experimented with a model trained on the data contained in this code.
\\\noindent\textbf{Baseline D}~\cite{DBLP:journals/corr/abs-2203-10408} extracts features from email headers and classifies them using a random forest classifier. It captures five categories of features, including count-based features of header fields and comparison-based features, such as matching domain names in header fields. We used a model trained on two datasets cited in the paper (TREC Public Spam Corpus and PhishingCorpus).

We describe the results of our experiments with these baseline systems. When compared to our system using GPT-4, it significantly outperformed all the baseline systems. Although baselines A to C can detect phishing emails with high accuracy when validated with test data split from the datasets mentioned in their respective papers or source code, they show accuracies of 60.00\%, 57.26\%, and 54.53\%, respectively, on our dataset. These systems demonstrated lower performance as they were trained specifically on the language, vocabulary, and context of their datasets, leading to reduced generalization capability across other types of data.
On the other hand, baseline D, which uniquely sets feature extraction from both the email header and body, could flexibly detect phishing emails compared to other baseline systems, achieving a moderate accuracy of 86.22\%. Our system surpasses the performance of baseline systems through a comprehensive approach that includes analyzing both the email body and header. Additionally, it leverages advanced contextual interpretation and incorporates pre-trained knowledge, such as legitimate domain names associated with brands, to effectively identify phishing emails.

\subsection{Examples of Detection Results}

\begin{figure}[!t]
    \centering
    \begin{subfigure}{0.45\linewidth}
        \centering
        \fbox{\includegraphics[width=0.95\linewidth]{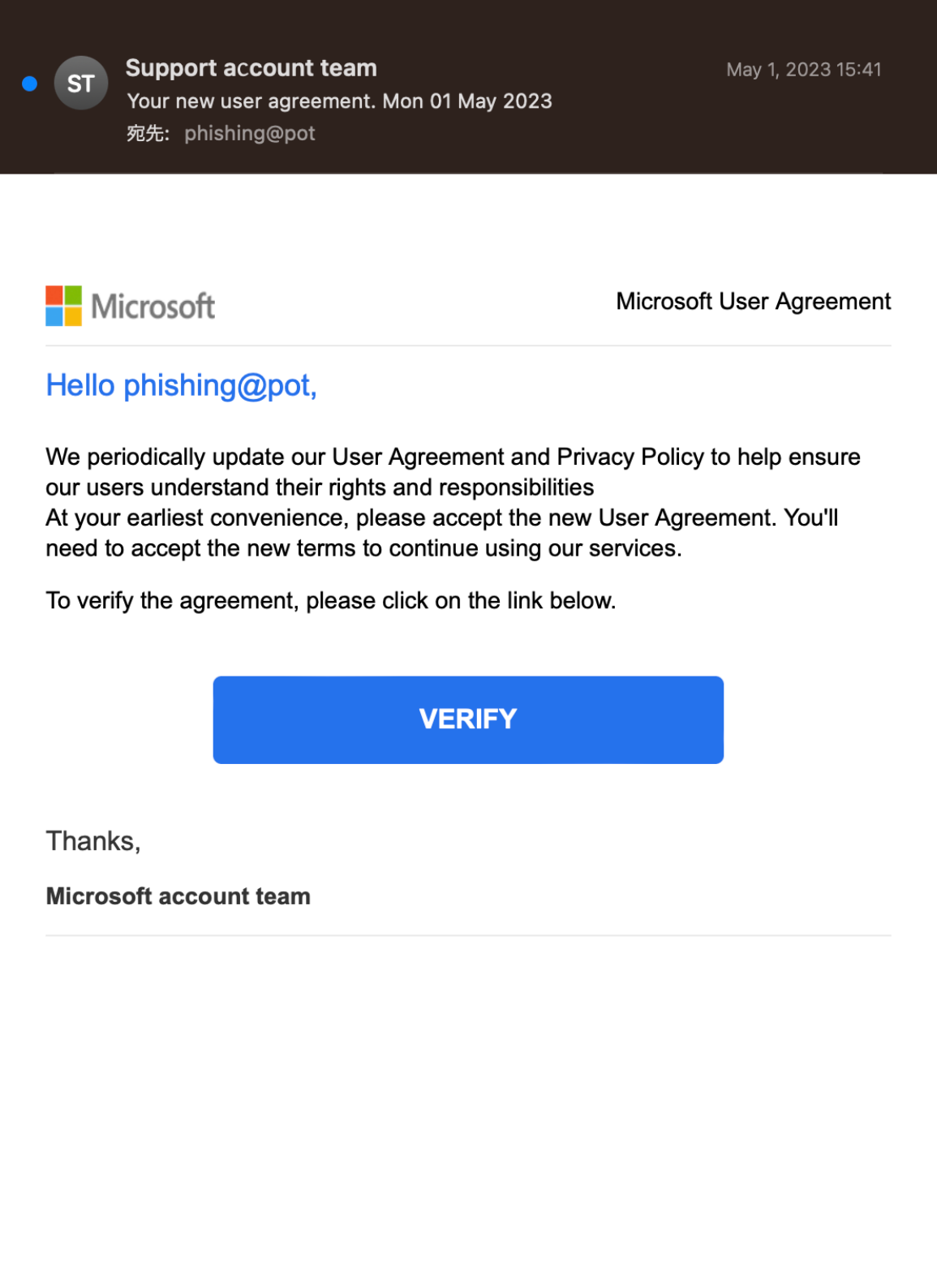}}
        \caption{Example A.}
        \label{fig:example-a}
    \end{subfigure}
    \begin{subfigure}{0.45\linewidth}
        \centering
        \fbox{\includegraphics[width=0.95\linewidth]{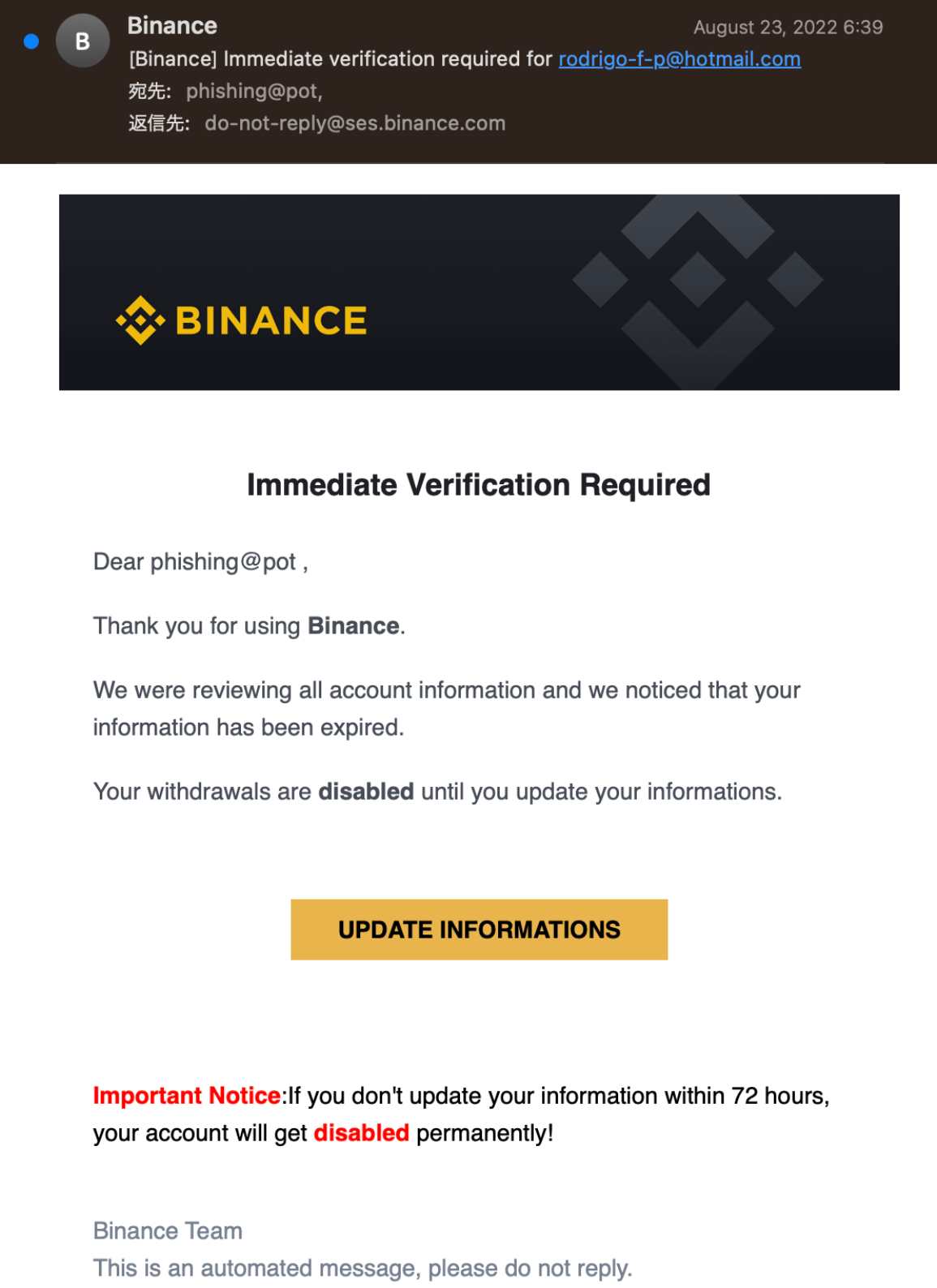}}
        \caption{Example B.}
        \label{fig:example-b}
    \end{subfigure}
    \caption{Examples of Phishing Emails}
    \label{fig:examples}
\end{figure}

\newtcbtheorem[]{response}{Response}%
{
  colback=responsecolor!10!white,
  colframe=responsecolor,
  coltext=black,
  boxsep=0pt,
  fonttitle=\bfseries,
  fontupper=\ttfamily,
  sharp corners
}{prompt}

\begin{figure}[!t]
\begin{response}[label=response1]{Example A}{response1}
\{\\
"is\_phishing": true,
\\"phishing\_score": 8,
\\"brand\_impersonated": "Microsoft",
\\"rationales": "This email appears to be a phishing attempt. The sender is ostensibly from 'crous-montpellier.fr', but the email's body text mentions the 'Microsoft account team', which is an impersonation of the Microsoft brand. This discrepancy is a sign of phishing. The email subject line, 'Your new user agreement', is designed to create a sense of urgency and lure the recipient into clicking on the link provided. The hyperlink uses a tracking ID which suggests that it might be used for collecting sensitive information. Examining the authentication results from the header, the emails passed SPF (Sender Policy Framework) and DKIM (DomainKeys Identified Mail) checks. However, this could still be sophisticated phishing, as attackers nowadays often compromise legitimate servers to launch phishing campaigns.",
\\"brief\_reason": "Impersonation of a well-known brand (Microsoft), inconsistency in sender's information, and presence of a suspicious hyperlink."
\\\}
\end{response}
\end{figure}

\begin{figure}[!t]
\begin{response}[label=response2]{Example B}{response2}
\{
\\"is\_phishing": true,
\\"phishing\_score": 9,
\\"brand\_impersonated": "Binance",
\\"rationales": "Several indicators lean towards this being a phishing email. To begin with, the sender's email address has a discrepancy: the email claims to be from 'Binance', but the sender's domain is not associated with Binance (ilonasavola.com). Moreover, the SPF check has failed as indicated in the 'Authentication-Results' field. The content of mail also raises suspicion: it impersonates Binance, creates a sense of urgency by stating that withdrawals are disabled and the account will be permanently disabled without immediate action and uses a disguised link for 'update informations', which leads to a different domain (zzdzw.com). These elements are common tactics employed in phishing emails, designed to deceive users into providing personal information.",
\\"brief\_reason": "Spoofed sender address and domain, impersonation of Binance, sense of urgency, disguised potentially harmful link, and SPF check failure."
\\\}
\end{response}
\end{figure}

In this section, we present examples of detection results of phishing emails. Figure~\ref{fig:example-a} shows an example of a phishing email from our dataset. This email pretends to be from Microsoft and asks the recipient to accept an update to the user agreement by clicking a "verify" button that leads to a phishing site. The domain name linked from this button has been flagged as phishing by three vendors on VirusTotal~\cite{virustotal}. 
The analysis results by GPT-4 are shown in Response~\ref{response1}, where the \texttt{is\_phishing} flag is \texttt{true}, and the \texttt{phishing\_score} is 8, indicating successful phishing detection with a high score (out of 10). 
As a basis for its determination, GPT-4 points out that in the body of the email, it mentions ``Microsoft Account Team,'' yet the sender's email domain name is not associated with Microsoft.
In addition, GPT-4 found authentication passes by analyzing the headers, but still correctly identified the email as potentially phishing.

Example B, shown in Figure~\ref{fig:example-b}, is a phishing email impersonating the cryptocurrency company Binance, warning that account information has expired and urging an update of account details. A link in the button leads to a URL unrelated to Binance. GPT-4's analysis results are shown in Response~\ref{response2}, where the \texttt{is\_phishing} flag is \texttt{true} and receives a high \texttt{phishing\_score} of 9. The \texttt{brand\_impersonated} is correctly identified as Binance. Despite the \texttt{From} header containing \texttt{Binance <do-not-reply@ses.binance.com>}, which appears to be the legitimate display name, GPT-4 identifies the actual sender as unrelated to Binance due to the failure of the SPF check. It also identifies the use of a ``sense of urgency'', a common SE technique, and the redirection to another domain name, correctly identifying it as an attempt to steal personal information.

\subsection{LLM Capabilities for Phishing Email Detection}
In this section, we investigated the strengths of the proposed system using LLMs by conducting a detailed analysis of their responses. Our results highlight the superior analytical capabilities of our system in four key factors.

\subsubsection{Extracting and Comparing Information Contained in Headers}
LLMs perform advanced analysis of email headers, extracting critical information primarily from the sender address, authentication results (\texttt{pass}/\texttt{fail}), subject line, and communication paths. The sender's address is the most important factor in analyzing the authenticity of an email. Although the From header is easily spoofed, LLMs can determine the legitimacy of an email by combining the address with the authentication results. 
In addition, LLMs check for discrepancies between the sender's domain name and the originating server in the Received header. 
For example, GPT-4 flagged a phishing email purportedly from \texttt{Trust Wallet <noreply@support-trustwallet[.]com>} as a spoofing attempt based on two critical observations: the domain name did not appear in the \texttt{Received} line, and the \texttt{Authentication-Results} field indicated a \texttt{softfail} for both SPF and DMARC checks.
LLMs also identify SE techniques from the subject line, a key factor in enticing users to open an email, including psychological tactics to create curiosity or a sense of urgency.

\urlstyle{tt}
\subsubsection{Identifying SE Techniques in the Body}
The body of an email often contains various SE techniques aimed at encouraging users to click on links. Phishing emails typically exploit the trust and recognition of well-known brands to lower users' caution and encourage them to follow the instructions. Attackers try to deceive recipients by constructing convincing scenarios, such as ``payment confirmation'' posing as online banking or ``security issues with your account'' posing as a social media platform. In these cases, LLMs can accurately identify the brand impersonation, not merely through Named Entity Recognition (NER) but by analyzing the context to determine which brand is being impersonated.

For example, GPT-4 identified that a phishing email, with the display name \texttt{service@lnt.\includegraphics[height=0.9\baselineskip]{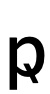}aypal[.]com} (where the first ``p'' is a Cyrillic character) and ``P$\alpha$yPal'' (where the first ``a'' is represented by the alpha) in the body, uses non-standard characters to bypass spam filters. GPT-4 correctly specified the brand and SE techniques to create a sense of urgency, such as ``our systems have detected violated terms \& policy.''
As another example, GPT-4 correctly identified evasion techniques in a phishing emails, where \includegraphics[height=0.9\baselineskip]{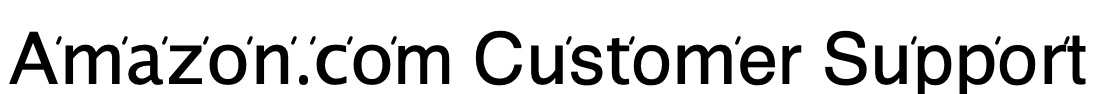} (with all characters separated by U+070F ``Syriac Abbreviation Mark'') appeared in the both \texttt{From} field and body. 
Additionally, GPT-4 successfully identified a disguised hyperlink supposedly leading to \urlstyle{tt}\url{https://www.amazon[.]com/}. It revealed that all characters in the URL were also separated by the U+070F mark and the actual destination of the link was a Google Apps Script (\url{script.google[.]com}), indicating potentially malicious intent.

\subsubsection{Combining Multiple Clues to Identify Evidence}
LLMs can identify evidence leading to the determination of phishing emails by combining clues extracted from both headers and the body. Equipped with knowledge of brands and their associated legitimate domain names, LLMs can distinguish mismatches between these and the sender's address or URLs in the email body. 
Additionally, LLMs accurately recognize the use of domain names similar to legitimate ones or URL shortening services (e.g., \url{tinyurl[.]com} and \url{cutt[.]ly}) to redirect users to phishing sites, with the intention of evading spam filters. 
For example, regarding a phishing email with the sender \texttt{BÍôčkchaiň.čom <notify@gotgel[.]org>}, GPT-4 identified it as phishing under the \urlstyle{same}\url{Blockchain.com} brand and correctly identified the tactic, stating, ``The sender's use of broken language and special characters to replace normal alphabets shows attempts at bypassing spam filters.'' 
Furthermore, GPT-4 detected 124 phishing emails without impersonating brands, through the presence of SE techniques.
\urlstyle{tt}

\subsubsection{Generating Analysis Reports}
Our system enables LLMs to output both detailed rationales and simplified explanations, allowing users to understand why an analyzed email is suspected of being phishing or is considered legitimate. This capability supports users in making informed decisions about emails diverted to the spam folder, determining whether to ignore them as phishing attempts or to reconsider them as false positives.

\subsection{False Positive and Negative}
We focus on cases where our system made incorrect classifications by GPT-4 and GPT-3.5 to analyze the capabilities and limitations of the system.

\subsubsection{False Positive}
We analyzed false positives, where our system incorrectly identified legitimate emails as phishing.  Our system using GPT-4 produced only three false positives. One case involved a legitimate email from Tesco where the domain name of the sender address was different from the legitimate domain name and there were multiple subdomains, leading to misclassification. Another case involved a legitimate email from CNET where the complexity of the local part of the sender's address led to misclassification, compounded by the presence of multiple links in the email body, which raised suspicion. The third case was a promotional email for a new publication issue; although our system did not identify the brand, phrases such as "GET YOUR FREE NECKLACE NOW" led to misclassification as phishing because they appeared to be SE techniques.

On the other hand, GPT-3.5 had 32 false positives. These included misinterpreting SE techniques, failing to determine the legitimacy of the sender's address, and classifying phishing without evidence. Misinterpretations of SE techniques included responding to words such as ``free'' or ``discount'' and sensational content in news articles, such as corporate scandals. GPT-3.5 considered these to be common tactics used to attract users. 
In contrast, GPT-4 not only focuses on such emotional phrases but also performs a deep analysis to determine the correct SE technique based on how it leads to the phishing site.
GPT-3.5 also failed to correctly identify the legitimacy of the sender's address, for example, misclassifying an email from a user's Hotmail address as Hotmail phishing. In addition, there were cases where there was no evidence of phishing in the headers or body, yet the final classification was phishing. GPT-4 provided more detailed descriptions of the reasons compared to GPT-3.5. While GPT-4 detailed specific SE techniques and suspicious indicators in the headers, GPT-3.5 often made general statements about the presence of phishing indicators in the body or hyperlinks, or the presence of brand impersonation, without providing concrete evidence.

\subsubsection{False Negative}
We present cases in which the proposed system erroneously classified a phishing email as legitimate. Our system using GPT-4 had only three false negatives. Two of these were phishing emails claiming to be from FTX, both attempting to trick recipients into submitting customer claims by a certain deadline. The system missed the phishing email because the authentication of the domain name was successful and the link in the body contained a legitimate FTX domain name. In another case, an email from Tether Cashback asked the recipient to click a link to activate a new account. The system determined that this was not phishing because it found no spoofing or false information in the email.

In comparison, GPT-3.5 had 30 false negatives. In some cases, no phishing evidence was detected at all, while in others, some evidence was detected but the email was misclassified as legitimate when combined with other evidence. 
As an example of the former, a phishing site posed as Microsoft Privacy, requesting an ``update to the latest User Agreement.'' It exploited Bing's redirect feature with a URL starting with \url{https://www.bing[.]com/ck/a?!&&p=}, leading to a phishing site.
GPT-3.5 correctly recognized the Microsoft brand but mistakenly deemed it legitimate due to the recipient's server being \url{outlook[.]com}, despite the sender's address being unrelated to Microsoft. 
On the other hand, GPT-4 correctly identified the phishing email by highlighting the SE techniques of urging users, along with the sender's address domain name discrepancy. It stated, ``The link provided in the email is obfuscated, potentially leading to a malicious site, as it does not direct to a Microsoft domain, instead using \url{bing[.]com}''.

One notable case involves a phishing email that abused the Google Document sharing feature and was sent from \urlstyle{tt}\url{Google Notifications <notify-noreply@google[.]com>}. GPT-3.5 detected nothing suspicious in the sender or body, as both were legitimate. However, we confirmed that accessing the Google Form link in the email led to a BTC withdrawal scam site. GPT-4 identified that the content and subject of the email used SE techniques to capture the user's interest. Surprisingly, it correctly recognized the email as phishing by identifying the common attacker tactic of using Google Forms to direct users to phishing sites. This ability to correctly identify phishing, even in emails that use legitimate services and contain links with legitimate domain names, is an example of GPT-4's advanced analysis capabilities.
Another example is an email with the subject line ``Update:Your wallet has failed the Merge'' pretending to be from the MetaMask cryptocurrency software wallet. GPT-3.5 noted that the sender address did not belong to the legitimate MetaMask domain name and that the body of the email contained suspicious elements urging the user to take action, but concluded with ``Further investigation is required,'' thus avoiding a definitive phishing judgment. In contrast, GPT-4 correctly identified these tactics and determined that the email was phishing.

\section{Related Work}
This section reviews the current state of research in phishing email detection, highlighting key methodologies and findings relevant to our work.

\noindent\textbf{Machine Learning and Deep Learning Approaches}
A wide range of research has been conducted on the detection of phishing emails, including proposals for effective detection methods, analysis of current security features, introduction of new protocols, and exploration of the human factors involved in phishing and their countermeasures. One approach in detecting phishing emails is using machine learning-based techniques, which involve extracting features from email headers and bodies and classifying them using various classifiers~\cite{DBLP:journals/cluster/MughaidAHTAE22,DBLP:conf/raid/NabeelASKWY21,DBLP:journals/sncs/Sonowal20}.
Additionally, deep learning-based natural language processing models have been used to interpret context and detect phishing tendencies. For instance, Li et al.~\cite{DBLP:journals/tbd/LiCWS22} employed Long Short-Term Memory (LSTM) networks for phishing detection in large-scale email data. Che et al.~\cite{DBLP:conf/qrs/CheLZYZY17} used a combination of semantic web databases and fuzzy logic control in their approach. Several studies have proposed using BERT to analyze context for phishing email detection~\cite{DBLP:conf/eurosp/LeeTYAHD21,DBLP:conf/codaspy/QachfarVM22,DBLP:conf/ccs/SakaVK22}. Our system aims to conduct more advanced analyses by using LLMs to analyze the context of emails, allowing for a more comprehensive analysis of SE techniques present in both the body and headers.
Heiding et al.~\cite{Heiding2024DevisingAD} explored the use of LLMs for generating and detecting the intention behind phishing emails. They compared the success rates of LLM-generated phishing emails with those created by human experts and evaluate the ability of LLMs to detect malicious intent, highlighting both the promising results and potential for misuse.
While their study focused on detecting malicious intent, our research provides a detailed analysis of the phishing email detection capabilities of LLMs by creating a dataset containing both phishing emails and legitimate emails.
Chataut et al.~\cite{DBLP:conf/ccwc/ChatautGU24} evaluated the ability of LLMs to identify phishing attempts in email bodies.
Si et al.~\cite{si2024evaluatingperformancechatgptspam} developed a Chinese phishing email dataset and proposed an in-context learning approach for LLMs to classify phishing emails. Their research showed that increasing the number of few-shot examples improved classification accuracy.
While these studies primarily focused on analyzing email bodies to identify phishing attempts, our system takes a more comprehensive approach. We use LLMs to analyze all email components, including headers. This enables the detection of various phishing techniques, such as sender spoofing, and provides a more robust identification system.

\noindent\textbf{Summarization and User Interpretation}
Kashapov et al. proposed a method to summarize emails, assisting users in identifying phishing emails~\cite{DBLP:conf/asiaccs/KashapovWAR22}. While their approach aligns with ours in providing human-interpretable information, our system further enhances user interpretation by incorporating detailed phishing detection results and their rationale. Comprehensive comparative evaluations of tools for phishing email analysis have been conducted in several studies~\cite{DBLP:conf/ndss/CranorEHZ07,DBLP:conf/soups/ZhengB23}. Methods for analyzing new types of phishing attacks and campaigns, such as Lateral Phishing~\cite{DBLP:conf/uss/CidonGBKST19} and Business Email Compromise~\cite{DBLP:conf/uss/HoCGSPSV019}, have been proposed and extensively investigated~\cite{DBLP:conf/raid/GasconUSR18,DBLP:journals/tdsc/GutierrezKCAGCB18,DBLP:conf/sac/HanS16,DBLP:conf/uss/Hu018,DBLP:conf/imc/SimoiuZTB20}.

\noindent\textbf{Security Features and Human Factors}
Research also exists on validating existing security features and testing new security functionalities and protocols~\cite{DBLP:conf/soups/0001SBHVSM23,DBLP:conf/uss/ShenWGZLLZHDP021,DBLP:conf/uss/StringhiniEZHKV12}. Various studies not only systematically analyze phishing emails but also focus on human factors to discuss reducing the damage caused by phishing emails. For example, research has been conducted on user behavior-focused countermeasures against phishing emails~\cite{pilavakis2023didn,DBLP:conf/soups/ZhengB22} and analyses of why users fall for phishing emails~\cite{greene2018user,DBLP:conf/uss/HeijdenA19,DBLP:journals/tissec/ZhuoBKLR23}. Additionally, there are studies on educating users and corporate training to prevent phishing emails~\cite{DBLP:conf/uss/BrunkenBHS23,jayakrishnan2022pickmail,DBLP:conf/soups/ReinheimerAMMDL20}.

\section{Discussion}
We discuss the limitations of our system and evaluation experiments.

\noindent\textbf{Scope of Phishing Emails}
In this study, we excluded phishing emails that do not contain links in their body from our scope. Examples of such phishing emails include those that attach PDF or Office format files containing malicious code or macros. It is challenging to directly determine whether an attachment's binary data is malware using LLMs. However, there is potential for phishing detection based on SE techniques described in the email body, or brand impersonation, suggesting that the overall email can be identified as phishing.
Furthermore, emails that do not lead users to phishing sites but instead request personal information or direct financial transfers (such as bank transfers or cryptocurrency) in response to the phishing email were also outside the scope of this research. Similarly, these emails could potentially be detected by our proposed system if they contain impersonation or deception, based on the context.

\noindent\textbf{LLM Parameters} 
In our research, we used the default parameters for each LLM, but adjusting these parameters could change the results.
For instance, Temperature and Top\_p are known to significantly influence the output. Temperature is a parameter that controls the randomness of the generated text, with higher values causing the model to produce more diverse and unpredictable text. Similarly, Top\_p is a parameter that regulates the diversity of the generated text, where higher values prompt the model to select from a broader range of high-probability tokens in the probability distribution.
We found that the detection accuracy of phishing emails was sufficiently high with the default parameters, and the rationale was thoroughly and concretely explained. However, if consistent answers are necessary or more flexible decision-making is desired, customizing the parameters could allow for adjustments tailored to specific needs.

\noindent\textbf{Improving Response Accuracy with RAG}
LLMs can enhance their outputs through Retrieval-Augmented Generation (RAG) by searching for information from external knowledge bases. 
In this study, we attempted to detect phishing emails using only the capabilities of LLMs without using such techniques, and we achieved high accuracy in distinguishing between phishing emails and legitimate emails.
This is because the LLMs we tested successfully identified the association between brands targeted by phishing emails and their legitimate domain names through pre-training on a variety of textual data.
Even for local brands in non-English regions, GPT-4 in particular was able to identify brand impersonation by comparing legitimate domain names.
However, LLMs cannot identify the existence of new brands that have emerged after the training period or newly acquired legitimate domain names.
This is where RAG may work effectively to overcome such limitations of LLMs.
For example, we can provide additional domain name knowledge to LLMs by creating a list of legitimate domain names or by integrating LLMs with search engines to provide real-time access to up-to-date domain information.
Such an extension of the capabilities of LLMs using RAG could be a future research topic.

\noindent\textbf{Deployment Scenarios}
We discuss the real-world deployment scenarios for our system.
Our evaluation experiments have shown that our system has high detection capabilities for multilingual phishing emails impersonating various brands.
For real-world deployment, the system could potentially replace existing phishing email defense mechanisms implemented in email services or analyze individual emails on user devices. However, as shown in section~\ref{section:results}, there is a cost associated with each email analysis. This means that analyzing all emails received by a company or organization using our system could potentially incur significant costs.
However, given the intense competition in LLM development, there is an ongoing trend of more powerful model APIs becoming available at lower prices. In fact, as of May 2024, the API costs for \texttt{gpt-4-turbo} have been significantly reduced compared to the \texttt{gpt-4-0613} model we used in our experiments, with input costs decreasing from \$0.03/1K tokens to \$0.01/1K tokens and output costs from \$0.06/1K tokens to \$0.03/1K tokens.
Furthermore, the development of high-performance open-source models is actively progressing, and in the future, the cost of analyzing large volumes of emails may no longer be a concern.
Realistic scenarios include using the proposed system for email analysis by security vendors to generate intelligence, or by end users to analyze suspicious emails that have been classified into spam folders or have reached their inbox, on a case-by-case basis.
Moreover, using lower-cost models like GPT-3.5, which achieved a high accuracy of 96.92\% in our experiments, for the initial analysis and then performing detailed analysis using more accurate models only for results with ambiguous scores could also be an effective approach.

\section{Conclusion}
In this paper, we proposed \textsc{ChatSpamDetector}, a novel system for detecting phishing emails. This system examines both the headers and the body of emails to identify various deceptive strategies, including brand impersonation and social engineering tactics. Moreover, \textsc{ChatSpamDetector} can provide detailed explanations for its determinations, drawing on specific evidence to confirm an email as a phishing attempt. 

Our evaluation experiments demonstrated that \textsc{ChatSpamDetector} significantly outperforms existing baseline systems, achieving a remarkable detection accuracy of 99.70\% by using GPT-4. Unlike existing spam filters that rely on continuous updates to their models and block lists—after analyzing a large corpus of phishing emails—our system excels at identifying a wide range of phishing emails across multiple languages with high accuracy, without necessitating further training. This system not only provides a new option for preventing phishing emails, but also enables users to make informed decisions by providing concrete rationales for the suspiciousness of emails.

\urlstyle{rm}
\bibliographystyle{splncs04}
\bibliography{bib}

\begin{thebibliography}{10}
\providecommand{\url}[1]{\texttt{#1}}
\providecommand{\urlprefix}{URL }
\providecommand{\doi}[1]{https://doi.org/#1}

\bibitem{treccorpus}
{2007 TREC Public Spam Corpus}. \url{https://plg.uwaterloo.ca/~gvcormac/treccorpus07/} (2007)

\bibitem{csdmc}
{CSDMC Spam Corpus} (2021), \url{https://csmining.org/cdmc2021/datasets/}

\bibitem{dbsheta}
{dbsheta/spam-detection-using-deep-learning}. \url{https://github.com/dbsheta/spam-detection-using-deep-learning} (2024)

\bibitem{enron}
{Enron Email Dataset}. \url{https://www.cs.cmu.edu/~enron/} (2024)

\bibitem{mo-messidi}
{mo-messidi/Email-Phishing-Attempts-Detection-using-NLP}. \url{https://github.com/mo-messidi/Email-Phishing-Attempts-Detection-using-NLP} (2024)

\bibitem{MoAbd}
{MoAbd/Spam-detection}. \url{https://github.com/MoAbd/Spam-detection} (2024)

\bibitem{phishingpot}
{rf-peixoto/phishing\_pot}. \url{https://github.com/rf-peixoto/phishing\_pot} (2024)

\bibitem{spamassasincorpus}
{SpamAssassin public mail corpus}. \url{https://spamassassin.apache.org/old/publiccorpus/} (2024)

\bibitem{virustotal}
{VirusTotal}. \url{https://www.virustotal.com/} (2024)

\bibitem{DBLP:journals/corr/abs-2203-10408}
Beaman, C., Isah, H.: Anomaly detection in emails using machine learning and header information. CoRR  \textbf{abs/2203.10408} (2022)

\bibitem{DBLP:conf/uss/BrunkenBHS23}
Brunken, L., Buckmann, A., Hielscher, J., Sasse, M.A.: "to do this properly, you need more resources": The hidden costs of introducing simulated phishing campaigns. In: Calandrino, J.A., Troncoso, C. (eds.) {USENIX} Security 2023 (2023)

\bibitem{DBLP:conf/ccwc/ChatautGU24}
Chataut, R., Gyawali, P.K., Usman, Y.: Can {AI} keep you safe? {A} study of large language models for phishing detection. In: Paul, R., Kundu, A. (eds.) 14th {IEEE} Annual Computing and Communication Workshop and Conference, {CCWC} 2024, Las Vegas, NV, USA, January 8-10, 2024. pp. 548--554. {IEEE} (2024). \doi{10.1109/CCWC60891.2024.10427626}, \url{https://doi.org/10.1109/CCWC60891.2024.10427626}

\bibitem{DBLP:conf/qrs/CheLZYZY17}
Che, H., Liu, Q., Zou, L., Yang, H., Zhou, D., Yu, F.: A content-based phishing email detection method. In: 2017 {IEEE} International Conference on Software Quality, Reliability and Security Companion (2017)

\bibitem{DBLP:conf/uss/CidonGBKST19}
Cidon, A., Gavish, L., Bleier, I., Korshun, N., Schweighauser, M., Tsitkin, A.: High precision detection of business email compromise. In: 28th {USENIX} Security Symposium (2019)

\bibitem{DBLP:conf/ndss/CranorEHZ07}
Cranor, L.F., Egelman, S., Hong, J.I., Zhang, Y.: Phinding phish: An evaluation of anti-phishing toolbars. In: {NDSS} 2007 (2007)

\bibitem{DBLP:conf/raid/GasconUSR18}
Gascon, H., Ullrich, S., Stritter, B., Rieck, K.: Reading between the lines: Content-agnostic detection of spear-phishing emails. In: Research in Attacks, Intrusions, and Defenses - 21st International Symposium, {RAID} 2018 (2018)

\bibitem{geminiapi}
{Google Cloud}: {Gemini API}. \url{https://cloud.google.com/vertex-ai/docs/generative-ai/model-reference/gemini} (2024)

\bibitem{gmail-spam}
{Google Workspace Blog}: {An overview of Gmail's spam filters}. \url{https://workspace.google.com/blog/identity-and-security/an-overview-of-gmails-spam-filters?hl=en} (2024)

\bibitem{greene2018user}
Greene, K.K., Steves, M., Theofanos, M., Kostick, J., et~al.: User context: an explanatory variable in phishing susceptibility. In: 2018 Workshop Usable Security (2018)

\bibitem{DBLP:journals/tdsc/GutierrezKCAGCB18}
Gutierrez, C.N., Kim, T., Corte, R.D., Avery, J., Goldwasser, D., Cinque, M., Bagchi, S.: Learning from the ones that got away: Detecting new forms of phishing attacks. {IEEE} Trans. Dependable Secur. Comput.  \textbf{15}(6),  988--1001 (2018)

\bibitem{DBLP:conf/sac/HanS16}
Han, Y., Shen, Y.: Accurate spear phishing campaign attribution and early detection. In: 31st Annual {ACM} Symposium on Applied Computing (2016)

\bibitem{Heiding2024DevisingAD}
Heiding, F., Schneier, B., Vishwanath, A., Bernstein, J., Park, P.S.: Devising and detecting phishing emails using large language models. IEEE Access  \textbf{12},  42131--42146 (2024)

\bibitem{DBLP:conf/uss/HeijdenA19}
Heijden, A., Allodi, L.: Cognitive triaging of phishing attacks. In: 28th {USENIX} Security Symposium (2019)

\bibitem{DBLP:conf/uss/HoCGSPSV019}
Ho, G., Cidon, A., Gavish, L., Schweighauser, M., Paxson, V., Savage, S., Voelker, G.M., Wagner, D.A.: Detecting and characterizing lateral phishing at scale. In: 28th {USENIX} Security Symposium (2019)

\bibitem{DBLP:conf/uss/Hu018}
Hu, H., Wang, G.: End-to-end measurements of email spoofing attacks. In: 27th {USENIX} Security Symposium (2018)

\bibitem{jayakrishnan2022pickmail}
Jayakrishnan, G., Banahatti, V., Lodha, S.: Pickmail: a serious game for email phishing awareness training. In: Usable Security and Privacy (USEC) Symposium (2022)

\bibitem{DBLP:conf/asiaccs/KashapovWAR22}
Kashapov, A., Wu, T., Abuadbba, S., Rudolph, C.: Email summarization to assist users in phishing identification. In: {ASIA} {CCS} '22 (2022)

\bibitem{koide2023detecting}
Koide, T., Fukushi, N., Nakano, H., Chiba, D.: Detecting phishing sites using chatgpt. CoRR  \textbf{abs/2306.05816} (2023)

\bibitem{KojimaGRMI22}
Kojima, T., Gu, S.S., Reid, M., Matsuo, Y., Iwasawa, Y.: Large language models are zero-shot reasoners. In: NeurIPS (2022)

\bibitem{rfc6376}
Kucherawy, M., Crocker, D., Hansen, T.: {DomainKeys Identified Mail (DKIM) Signatures}. RFC 6376 (Sep 2011)

\bibitem{rfc7489}
Kucherawy, M., Zwicky, E.: {Domain-based Message Authentication, Reporting, and Conformance (DMARC)}. RFC 7489 (Mar 2015)

\bibitem{DBLP:conf/eurosp/LeeTYAHD21}
Lee, J., Tang, F., Ye, P., Abbasi, F., Hay, P., Divakaran, D.M.: D-fence: {A} flexible, efficient, and comprehensive phishing email detection system. In: {IEEE} European Symposium on Security and Privacy, EuroS{\&}P (2021)

\bibitem{DBLP:journals/tbd/LiCWS22}
Li, Q., Cheng, M., Wang, J., Sun, B.: {LSTM} based phishing detection for big email data. {IEEE} Trans. Big Data  \textbf{8}(1),  278--288 (2022)

\bibitem{DBLP:conf/soups/0001SBHVSM23}
Liu, E., Sun, L., Bellon, A., Ho, G., Voelker, G.M., Savage, S., Munyaka, I.N.S.: Understanding the viability of gmail's origin indicator for identifying the sender. In: Nineteenth Symposium on Usable Privacy and Security, {SOUPS} 2023 (2023)

\bibitem{azureopenai}
{Microsoft Azure}: {Azure OpenAI Service}. \url{https://azure.microsoft.com/en-us/products/ai-services/openai-service} (2024)

\bibitem{outlook-spam}
{Microsoft Support}: {Overview of the Junk Email Filter}. \url{https://support.microsoft.com/en-us/office/overview-of-the-junk-email-filter-5ae3ea8e-cf41-4fa0-b02a-3b96e21de089} (2024)

\bibitem{DBLP:journals/cluster/MughaidAHTAE22}
Mughaid, A., AlZu'bi, S., Hnaif, A., Taamneh, S., Alnajjar, A., Elsoud, E.A.: An intelligent cyber security phishing detection system using deep learning techniques. Clust. Comput.  \textbf{25}(6),  3819--3828 (2022)

\bibitem{DBLP:conf/raid/NabeelASKWY21}
Nabeel, M., Altinisik, E., Sun, H., Khalil, I., Wang, W.H., Yu, T.: {CADUE:} content-agnostic detection of unwanted emails for enterprise security. In: {RAID} '21 (2021)

\bibitem{pilavakis2023didn}
Pilavakis, N., Jenkins, A., K{\"o}kciyan, N., Vaniea, K.: “i didn’t click”: What users say when reporting phishing. In: USEC 2023 (2023)

\bibitem{DBLP:conf/codaspy/QachfarVM22}
Qachfar, F.Z., Verma, R.M., Mukherjee, A.: Leveraging synthetic data and {PU} learning for phishing email detection. In: {CODASPY} '22: Twelveth {ACM} Conference on Data and Application Security and Privacy (2022)

\bibitem{DBLP:conf/soups/ReinheimerAMMDL20}
Reinheimer, B., Aldag, L., Mayer, P., Mossano, M., Duezguen, R., Lofthouse, B., von Landesberger, T., Volkamer, M.: An investigation of phishing awareness and education over time: When and how to best remind users. In: {SOUPS} 2020 (2020)

\bibitem{DBLP:conf/ccs/SakaVK22}
Saka, T., Vaniea, K., K{\"{o}}kciyan, N.: Context-based clustering to mitigate phishing attacks. In: AISec 2022 (2022)

\bibitem{rfc4408}
Schlitt, W., Wong, M.W.: {Sender Policy Framework (SPF) for Authorizing Use of Domains in E-Mail, Version 1}. RFC 4408 (Apr 2006)

\bibitem{brand-indicators-for-message-identification-04}
{Seth Blank et al.}: {Brand Indicators for Message Identification (BIMI)}. Internet-draft (2023), \url{https://datatracker.ietf.org/doc/draft-brand-indicators-for-message-identification/04/}, work in Progress

\bibitem{DBLP:conf/uss/ShenWGZLLZHDP021}
Shen, K., Wang, C., Guo, M., Zheng, X., Lu, C., Liu, B., Zhao, Y., Hao, S., Duan, H., Pan, Q., Yang, M.: Weak links in authentication chains: {A} large-scale analysis of email sender spoofing attacks. In: 30th {USENIX} Security Symposium (2021)

\bibitem{si2024evaluatingperformancechatgptspam}
Si, S., Wu, Y., Tang, L., Zhang, Y., Wosik, J.: Evaluating the performance of chatgpt for spam email detection (2024), \url{https://arxiv.org/abs/2402.15537}

\bibitem{DBLP:conf/imc/SimoiuZTB20}
Simoiu, C., Zand, A., Thomas, K., Bursztein, E.: Who is targeted by email-based phishing and malware?: Measuring factors that differentiate risk. In: {IMC} '20 (2020)

\bibitem{DBLP:journals/sncs/Sonowal20}
Sonowal, G.: Phishing email detection based on binary search feature selection. {SN} Comput. Sci.  \textbf{1}(4), ~191 (2020)

\bibitem{DBLP:conf/uss/StringhiniEZHKV12}
Stringhini, G., Egele, M., Zarras, A., Holz, T., Kruegel, C., Vigna, G.: B@bel: Leveraging email delivery for spam mitigation. In: 21th {USENIX} Security Symposium (2012)

\bibitem{Wei0SBIXCLZ22}
Wei, J., Wang, X., Schuurmans, D., Bosma, M., Ichter, B., Xia, F., Chi, E.H., Le, Q.V., Zhou, D.: Chain-of-thought prompting elicits reasoning in large language models. In: NeurIPS (2022)

\bibitem{white2023prompt}
White, J., Fu, Q., Hays, S., Sandborn, M., Olea, C., Gilbert, H., Elnashar, A., Spencer{-}Smith, J., Schmidt, D.C.: A prompt pattern catalog to enhance prompt engineering with chatgpt. CoRR  \textbf{abs/2302.11382} (2023)

\bibitem{DBLP:conf/pam/YajimaCYM23}
Yajima, M., Chiba, D., Yoneya, Y., Mori, T.: A first look at brand indicators for message identification {(BIMI)}. In: {PAM} 2023 (2023)

\bibitem{DBLP:conf/soups/ZhengB22}
Zheng, S.Y., Becker, I.: Presenting suspicious details in user-facing e-mail headers does not improve phishing detection. In: {SOUPS} 2022 (2022)

\bibitem{DBLP:conf/soups/ZhengB23}
Zheng, S.Y., Becker, I.: Checking, nudging or scoring? evaluating e-mail user security tools. In: {SOUPS} 2023 (2023)

\bibitem{DBLP:journals/tissec/ZhuoBKLR23}
Zhuo, S., Biddle, R., Koh, Y.S., Lottridge, D.M., Russello, G.: Sok: Human-centered phishing susceptibility. {ACM} Trans. Priv. Secur.  \textbf{26}(3),  24:1--24:27 (2023)

\end{thebibliography}
\end{document}